\documentclass[conference]{IEEEtran}
\usepackage{graphicx}    
\usepackage{subcaption}  
\IEEEoverridecommandlockouts

\usepackage{cite}
\usepackage{amsmath,amssymb,amsfonts}
\usepackage{algorithmic}
\usepackage{graphicx}
\usepackage{textcomp}
\usepackage{xcolor}
\def\BibTeX{{\rm B\kern-.05em{\sc i\kern-.025em b}\kern-.08em
    T\kern-.1667em\lower.7ex\hbox{E}\kern-.125emX}}
\begin{document}

\title{A Model Aware AIGC Task Offloading Algorithm in IIoT Edge Computing\\
\thanks{This work was supported by the National Natural Science Foundation of China under Grants (U22A2054). }
}
\author{
    \IEEEauthorblockN{1\textsuperscript{st} Xin Wang}
    \IEEEauthorblockA{\textit{Guangxi University } \\
    \textit{Key Laboratory of Intelligent} \\
    \textit{Networking and Scenario System} \\
    \textit{(School of Information and} \\
    \textit{Communication, Guilin University} \\
    \textit{ of Electronic Technology)} \\
        Guilin, 541004, China \\
        23022201024@mails.guet.edu.cn}
    \and
    \IEEEauthorblockN{2\textsuperscript{nd} Xiaohuan Li}
    \IEEEauthorblockA{\textit{Guangxi University } \\
    \textit{Key Laboratory of Intelligent} \\
    \textit{Networking and Scenario System} \\
    \textit{(School of Information and} \\
    \textit{Communication, Guilin University} \\
    \textit{ of Electronic Technology)} \\
        Guilin, 541004, China \\
        lxhguet@guet.edu.cn\\
        0000-0001-9097-4236}
    \and
    \IEEEauthorblockN{3\textsuperscript{ed} Xun Wang $^{\star}$}
    \IEEEauthorblockA{\textit{Guangxi University } \\
    \textit{Key Laboratory of Intelligent} \\
    \textit{Networking and Scenario System} \\
    \textit{(School of Information and} \\
    \textit{Communication, Guilin University} \\
    \textit{ of Electronic Technology)} \\
        Guilin, 541004, China \\
        wangxun8511@163.com\\
        0000-0001-9545-7350}
}
\maketitle
\begin{abstract}
The integration of the Industrial Internet of Things (IIoT) with Artificial Intelligence-Generated Content (AIGC) offers new opportunities for smart manufacturing, but it also introduces challenges related to computation-intensive tasks and low-latency demands. Traditional generative models based on cloud computing are difficult to meet the real-time requirements of AIGC tasks in IIoT environments, and edge computing can effectively reduce latency through task offloading. However, the dynamic nature of AIGC tasks, model switching delays, and resource constraints impose higher demands on edge computing environments. To address these challenges, this paper proposes an AIGC task offloading framework tailored for IIoT edge computing environments, considering the latency and energy consumption caused by AIGC model switching for the first time. IIoT devices acted as multi-agent collaboratively offload their dynamic AIGC tasks to the most appropriate edge servers deployed with different generative models. A model aware AIGC task offloading algorithm based on Multi-Agent Deep Deterministic Policy Gradient (MADDPG-MATO) is devised to minimize the latency and energy. Experimental results show that MADDPG-MATO outperforms baseline algorithms, achieving an average reduction of 6.98\% in latency, 7.12\% in energy consumption, and a 3.72\% increase in task completion rate across four sets of experiments with model numbers ranging from 3 to 6, it is demonstrated that the proposed algorithm is robust and efficient in dynamic, high-load IIoT environments. 

\end{abstract}


\section{Introduction}
In recent years, Artificial Intelligence-Generated Content (AIGC) has developed rapidly and deeply integrated with the Industrial Internet of Things (IIoT), injecting new vitality into smart manufacturing \cite{1xu2018industry}\cite{2nakagawa2021industry}\cite{4leng2024unlocking}. Leveraging vast datasets for training, AIGC crafts high-quality content, empowering seamless intelligent decision-making and automation across industrial environments\cite{6rafique2020complementing}\cite{7davis2012smart}. However, AIGC was traditionally deployed in the cloud center (CC), offering powerful computing capabilities but faces challenges due to its computationally intensive nature\cite{8bi2014internet}. When cloud-based AIGC deployment processes massive data and performs complex model inference, it leads to high communication latency and heavy loads\cite{9wang2024generative}. In IIoT environments, many tasks such as AIGC workloads are latency-sensitive and require rapid processing, making low latency a critical factor, while cloud computing is unsuitable due to its limitations in real-time processing and scalability\cite{10xu2024unleashing}\cite{11guo2024large}. Edge computing effectively reduces latency and alleviates cloud pressure by offloading tasks to edge server (ES) near end device (ED), providing a viable solution for AIGC applications in IIoT\cite{13liu2024towards}\cite{14liu2025qos}\cite{15du2024diffusion}.

Although edge computing effectively addresses the latency and load issues of cloud-based AIGC deployment, enhancing its potential in smart manufacturing and IIoT, it still faces numerous challenges\cite{16akhlaqi2023task}\cite{17naouri2021novel}\cite{18xu2023joint}. First, task dynamism is prominent in AIGC, as requests are random and vary widely in model requirements. ES must flexibly adapt to this unpredictability and supporting diverse models, for example, diffusion models for predictive maintenance, GANs for production optimization, and anomaly detection models for safety monitoring\cite{19li2024ai}. Unlike the resource-rich cloud, ES with limited storage and computation cannot host all models simultaneously and incur increased latency and costs from downloading uncached models\cite{20chen2018data}. Furthermore, dynamic tasks require frequent model switching, involving parameter loading, network updates and inference initialization, which exacerbate loading and initialization delays\cite{21xiao2020system}. However, model absence and switching delays significantly affect performance, especially when large models produce high-quality content, making this issue increasingly severe\cite{22hevesli2024task}.

Existing research\cite{20chen2018data}\cite{21xiao2020system}\cite{22hevesli2024task}\cite{wang2024toward} had largely focused on optimizing traditional factors such as transmission latency, computational latency and scheduling latency, which indeed play a critical role in task processing and response time. However, as the complexity of IIoT and AIGC models increases, focusing solely on a traditional factors is insufficient to comprehensively enhance system performance. In \cite{19li2024ai}, Xu \textit{et al.} proposed high-quality content generated by large models often comes at the cost of increased latency, while the additional loading and initialization delays introduced by model switching have not been adequately addressed.

To tackle the challenges outlined above, we propose an optimized AIGC task offloading framework tailored for IIoT edge computing environments. The latency and energy consumption caused by AIGC model switching are considered in this framework and a model aware task offloading algorithm based on Multi-Agent Deep Deterministic Policy Gradient (MADDPG-MATO) is devised to minimize the latency and energy. The main contributions of this paper are summarized
as follows:
\begin{itemize}
	\item An AIGC tasks offloading framework in IIoT is constructed by incorporating generative model aware and matching into existing decentralized offloading framework. The latency and energy consumption of AIGC model switching are considered into the offloading strategy.
\end{itemize}
\begin{itemize}
	\item EDs in IIoT act as multiple reinforcement learning agents and collaboratively offload their dynamic AIGC tasks to the most appropriate ES with desired model. The optimal offloading decisions are obtained by MADDPG to minimize the latency and energy.
\end{itemize}
\section{System Model}
\subsection{System Framework}
We focus on offloading  AIGC tasks in smart IIoT and propose a decentralized offloading framework that considers model switching, as shown in Fig.1.

\begin{figure}[htbp]
	\centering{\includegraphics[width=0.9\columnwidth]{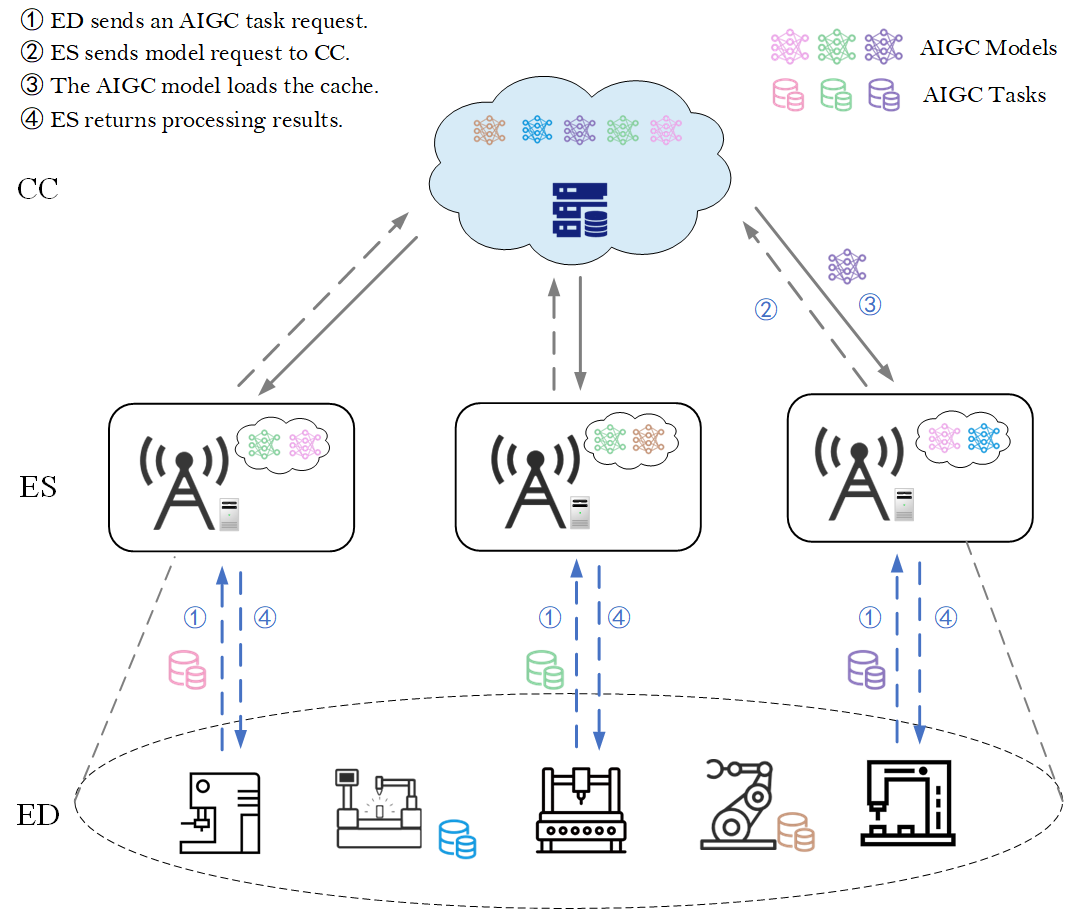}}
	\caption{System Model}
\end{figure} 
 AIGC models are typically trained in the CC, while inference is performed at the edge and terminal layers. In the framework, we define a set of ED as \( M = \{1, 2, \dots, M\} \) and a set of ES as \( N = \{1, 2, \dots, N\} \). Users can request AIGC tasks from any ED at any time, and these task requests are random. Therefore, we assume that the generation of AIGC tasks follows a uniform distribution and that task processing adheres to the First-In-First-Out (FIFO) principle. The ED sends a request for an AIGC task, and the algorithm determines the optimal decision by choosing whether to process it locally or offload it to the ES. If it is processed on the ES, model compatibility matching is required to check whether the ES has a model matching the task. If so, the task is processed directly; otherwise, the model is downloaded from the CC. After the ES completes the task, the results are returned to the ED.

The AIGC tasks primarily include: device health monitoring and prediction tasks, production process optimization tasks and industrial safety monitoring tasks. These tasks involve substantial data collection, real-time analysis, and computation, requiring significant computational resources and data transmission capacity. Once an AIGC task is initiated, the ED independently executes an offloading strategy based on its framework state, processing the task locally or offloading it to an ES. 

\subsection{Offloading Model}
This study aims to minimize the service latency and energy consumption of AIGC tasks and our goal is to address the following three key questions:

\begin{enumerate}
	\item Should an AIGC task be offloaded to an ES?
	\item How many tasks should be offloaded to the ES?
	\item Does this ES need to download the model for processing the task?
\end{enumerate}

Based on these considerations, we design the decision-making process for AIGC task offloading. We describe the task generated by ED \( m \) at time \( t \) as:
\begin{equation}
	l_{m, t}=\left\{\mu_{m, t}, x_{m, t}, \rho_{m, t}\right\},
\end{equation}
where \( \mu_m(t) \) denotes the type of AIGC task generated by ED \( m \) at time \( t \), \( x_m(t) \) represents the size of the AIGC task (in Mbits), and \( \rho_m(t) \) is the computational density (in cycles/bit).

We define a task offload target  $\xi_{m,t,n}\in \left\{0, 1\right\} $ to specify the ES for task offloading and an offloading rate $\eta_{m,t} \in [0, 1]$ to determine the portion of the AIGC task offloaded to the ES. 

We describe the definition of an AIGC model as:
\begin{equation}
	I = \left\{I_i, X_i\right\},
\end{equation}
where \( I_i \) is the unique index of the AIGC model, and \( X_i \) represents the size of the AIGC model (in Mbits). A binary download indicator \( \beta_{i,n} \in \{0, 1\} \) is defined for ES \( n \). Model compatibility $d_{m,i,n} \in \left\{0,1\right\}$ indicates whether ES \( n \) has the model required for task.

\subsection{Computation and Communication Model}
\subsubsection{Local Computation}
 The computation latency and energy consumption for local execution are given by:
\begin{equation}
T_{m,t}^{ed}(l) = \left\lceil \frac{x_{m,t} \cdot (1-\eta_{m,t}) \cdot  \rho_{m,t}}{f_m^{InDe}} \right\rceil,
\end{equation}
\begin{equation}
	E_{m,t}^{ed}(l)=c(f_m^{InDe}) \cdot x_{m,t} \cdot \rho_{\mathrm{m,t}}  ,
\end{equation} 
where \(x_{m,t}(1-\eta_{m,t}) \) represents the AIGC task data,  \( f_m^{InDe} \) denotes the computational capacity of the ED,  \( c \) is the effective switched capacitance of the ED's chip.

\subsubsection{Edge Server Computation}

When \( 0 < \eta_{m,t} \leq 1 \)) , the task will be offloaded to ES. Assuming the transmission rate is denoted as \( r_m^n \). The transmission latency and energy consumption for transferring a task from ED \( m \) to ES \( n \) are expressed as follows:
\begin{equation}
	T_{m,n}^{trans}(l)=\frac{x_{m,t}\eta_{m,t}}{r_m^n},
\end{equation}
\begin{equation}
	E_{m,n}^{trans}(l)=p_m^nT_{m,n}^{trans}(l),
\end{equation}
where \( p_m^n \) denotes the transmission power from ED \( m \) to the ES \( n \).

The model aware depends on \( \beta_{i,n} \) and \( d_{m,i,n} \): if the model is already cached, the switching latency is 0. If the model is not cached, it must be downloaded from the CC. Assum the transmission rate is denoted as \( r_c^n \). Based on these conditions, the model switching latency and energy consumption can be expressed as:
\begin{equation}
	T_{c,n}^{switch}(l)=\frac{X_i}{r_c^n},
\end{equation}
\begin{equation}
	E_{c,n}^{switch}(l)=p_c^nT_{c,n}^{switch}(l).
\end{equation}

\( f_n^{edge} \) is the computational capacity of ES \( n \). The computation latency and energy consumption of the task executed on the ES are respectively:

\begin{equation}
	T_n^{com}(l)=\left[\frac{x_{m,t}\cdot \eta_{m,t}\cdot \rho_{m,t}}{f_n^{edge}}\right],
\end{equation}
\begin{equation}
	E_n^{com}=c_n(f_n^{edge})^2 \cdot x_{m,t} \cdot \rho_{m,t}  .
\end{equation}

The total processing latency and energy consumption of the task on ES \( n \) are:
\begin{equation}
	T_{m,n}^{edge}(l)=T_{m,n}^{trans}(l)+T_{c,n}^{switch}(l)+T_n^{com}(l),
\end{equation}
\begin{equation}
	E_{m,n}^{edge}(l)=E_{m,n}^{trans}(l)+E_{c,n}^{switch}(l)+E_n^{com}(l).
\end{equation}

\subsection{Optimization Objective}
The objective of this paper is to minimize task latency and energy consumption. The completion latency and total energy consumption of a task are:
\begin{equation}
	T^{total}(l)=\max\left(T_{m,t}^{ed}(l),T_{m,n}^{edge}(l)\right),
\end{equation}
\begin{equation}
	E^{total}(l)=\max\left(E_{m,t}^{ed}(l),E_{m,n}^{edge}(l)\right).
\end{equation}

The optimization problem is formulated as:
\begin{equation}
	min\sum_{t=1}^{T}\sum_{l=1}^{L}\theta_{1}T^{total}(l)+\theta_{2}E^{total}(l) .\\
\end{equation}

Subject to:
$$\begin{aligned}
	& c_1{:} & & 1\leq m\leq M,1\leq n\leq N,\forall t\in T ,\\
	& c_2{:} & & \eta_{m,t}\in[0,1],\forall m\in M ,\\
	& c_3{:} & & \xi_{m,t,n}\in\{0,1\},\forall m\in M,\forall n\in N ,\\
	& c_4{:} & & f_{m}^{InDe}\leq F_{ED_{m}}^{Max},f_{n}^{edge}\leq F_{ES_{n}}^{Max},\forall m\in M,\forall n\in N ,\\
	& c_5{:} & & r_{m}^{n}\leq B_{n}^{max},\forall m\in M,\forall n\in N.
\end{aligned}$$

The $c_1$ governs task execution time and the number of EDs and ESs. The $c_2$ and $c_3$ ensure that AIGC tasks are processed on EDs or offloaded to ESs without exceeding the original data volume. $c_4$ caps task computational resources at the maximum capacities of EDs ($F_{ED_{m}}^{Max}$) and ESs ($F_{ES_{n}}^{Max}$). The $c_5$ limits task transmission rates to the communication capacity of each ES ($B_{n}^{max}$).

\section{Offloading Method Based on MADDPG}
\subsection{Markov Decision Process}
Since ED can be regarded as agents, the optimization problem can be modeled as a multi-agent Markov Decision Process (MDP) represented by the tuple \( < S, A, R, \gamma > \), where S, A, R respectively represent state, action space and reward of agents and \( \gamma \)  is the discount factor.
\subsubsection{State Space}
Each time, each ED node can obtain state information from the IIoT. It includes AIGC task type \( \mu_{m,t} \), size \( x_{m,t} \), computational density \( \rho_{m,t} \), edge server capacity \( f_n^{edge} \), model compatibility \( d_{m,i,n} \), and locations of EDs \( h_{EDm}(t) \), ESs \( h_{ESn}(t) \), and CC \( h_{CC}(t) \). The state space is expressed as:
\begin{equation}
	\begin{split}
		s_m(t) = \{ & \mu_{m,t}, x_{m,t}, \rho_{m,t}, f_n^{\text{edge}}, \\
		& d_{m,i,n}, h_{\text{ED}_m}(t), h_{\text{ES}_n}(t), h_{\text{CC}}(t) \}.
	\end{split}
\end{equation}

\subsubsection{Action Space}

The action space \( A \) consists of the actions of all agents, the action of agent \( a_m(t) \) includes the offloading target \( \xi_{m,t,n} \), offloading ratio \( \eta_{m,t} \), and whether to download the model \( \beta_{i,n} \). The action space is represented as:
\begin{equation}
	a_m(t)=
	\begin{Bmatrix}
		\xi_{m,t,n},\eta_{m,t},\beta_{i,n}
	\end{Bmatrix}.
    \end{equation}

\subsubsection{Reward Function}
The reward function aims to minimize latency and energy consumption, defined as:
\begin{equation}
	\begin{aligned}
		R_m\{s_m(t),a_m(t)\} &= \\ \sum_{l=1}^L\sum_{t=1}^T\sum_{m=1}^M\gamma^t& \cdot \begin{Bmatrix}-\sigma_l\omega_1T_m^{total}(t)-\sigma_l\omega_2E_m^{total}(t)-P_e
		\end{Bmatrix},
	\end{aligned}
\end{equation}
where \( \gamma\in \left(0,1 \right) \) is the discount factor; \( w_1 \) and \( w_2 \) are the weights for latency reward and energy consumption reward,  with \( w_1 + w_2=1 \); \( \sigma_l \) is the weight associated with the AIGC task type;  \( T_m^{total}(l) \) and \( E_m^{total}(l) \) are the task completion latency and energy consumption of agent \( m \), respectively; and \( P_e \) is the penalty for exceeding limits. 



\subsection{MADDPG-Based AIGC Task Offloading Algorithm}

This paper addresses the efficient offloading of AIGC tasks in edge computing environments by proposing MADDPG-MATO. The algorithm leverages the multi-agent reinforcement learning framework of MADDPG and trains the offloading strategy for each ED using an actor-critic network structure. Each ED agent uses state information to select task offloading targets and proportions and trigger model downloads, minimizing latency and energy while ensuring AIGC compatibility.

\subsubsection{Actor-Critic Network Structure}

Each ED agent \( m \) with an actor network and a critic network, which are used to generate actions and evaluate action values, respectively. 

\textbf{Actor}: The actor network \( v_m(o_m;\theta_m^v) \)  is responsible for generating agent actions. The input is the observation of the agent \( m \), denoted as \( o_m \), and the output is the action \( a_m \). 

\textbf{Critic}: The critic network \( Q_m(s;a;\theta_m^Q) \) evaluates the value of actions. The input consists of the global state \( s(t) \) and the actions of all agents \( A(t) \), and the output is the Q-value \( Q_m(s;a) \). 

\subsubsection{Update process}
First, randomly sample a batch of experiences from the replay buffer \( D \), and compute the target Q-value:
\begin{equation}
	y=R+\gamma Q’(s’,a;\theta^{Q’}).
\end{equation}


The critic network is updated by minimizing the loss function:
\begin{equation}
	L\left(\theta_m^Q\right)=\mathbb{E}_{s,a,R,s^{\prime}\thicksim D}\left[\left(y-Q_i(s,a;\theta_m^Q)\right)^2\right].
\end{equation}

The actor network is updated using the policy gradient:
\begin{equation}
	\begin{aligned}
		\nabla_{\theta_m^v}J(\theta_m^v) &= \mathbb{E}_{s\sim D}\Bigl[\nabla_{\theta_m^v}\nu_m(o_m;\theta_m^v) \\
		&\quad \cdot \nabla_{a_m}Q_m(s,a;\theta_m^Q)\big|_{a_m=v_m(o_m)}\Bigr].
	\end{aligned}
\end{equation}

Finally, to improve the stability of neural network training, the target network parameters of the actor and critic are softly updated using a constant \( \tau \):
\begin{equation}
	\theta_m^{v^{\prime}}\leftarrow\tau\theta_m^v+(1-\tau)\theta_m^{v^{\prime}}  ,
\end{equation}
\begin{equation}
	\theta_m^{Q^{\prime}}\leftarrow\tau\theta_m^Q+(1-\tau)\theta_m^{Q^{\prime}}  .
\end{equation}

\subsubsection{Action Execution}
After training is complete, the algorithm uses the trained actor network to generate offloading strategies, returning the reward.

\section{Results and Analysis}
\subsection{Experimental Setup and Application Scenario}
To validate the performance of the proposed MADDPG-MATO algorithm in AIGC task offloading scenarios, we design an experimental scenario comprising 1 CC, 3 ESs and 10 EDs within a rectangular area of \( 1 \, \text{km} \times 1 \, \text{km} \). The computational capacities of the CC, ES and ED are set to 40 GHz, 7 GHz, and a random value in \( [1, 3] \, \text{GHz} \).

AIGC tasks are generated by the EDs, the task data size is generated randomly within \( [2, 20] \, \text{MB} \) depending on the task type. The size of the AIGC model ranges from [90, 250] MB, with specific compatibility determined by the task type. For training parameters, the MADDPG-MATO algorithm uses a learning rate of 0.001, a discount factor of 0.95, a soft update rate of 0.01, an experience replay buffer capacity of 10,000, and a batch size of 1024.

To compare algorithm performance, we implemented the following baseline algorithms:
\begin{itemize}
	\item \textbf{MADDPG-NoModel}: A version of MADDPG-MATO without model aware. Agents can't detect ES model compatibility or download models, used to measure the value of model aware in AIGC task offloading.
	
	\item \textbf{SADDPG}: An offloading algorithm based on single-agent DDPG, where each ED independently optimizes its offloading strategy, ignoring multi-agent collaboration.
	
	\item \textbf{Random}: Randomly selects offloading targets and ratios without considering model compatibility.
	
	\item \textbf{Greedy}: Prioritizes the nearest compatible ES; if no compatible ES is available, tasks are executed locally with a fixed offloading ratio of 1.0.
\end{itemize}
\subsection{Convergence Analysis}
To evaluate the convergence of MADDPG-MATO, we conducted comparative experiments with baseline algorithms MADDPG-NoModel, and SADDPG, with results shown in Fig.2. 
\begin{figure}[htbp]
	\centering{\includegraphics[width=0.95\columnwidth]{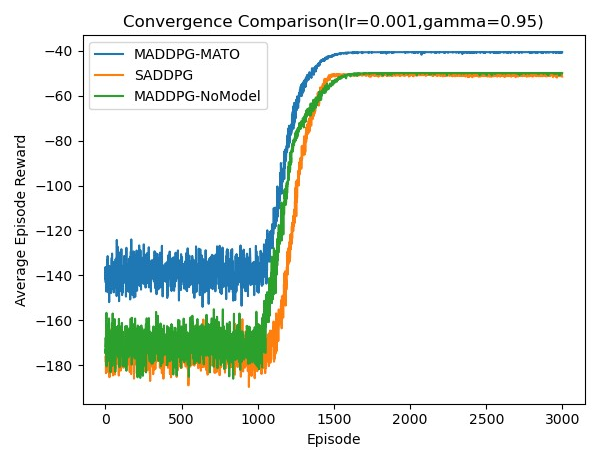}}
	\caption{Convergence Comparison}
\end{figure}

The figure illustrates that MADDPG-MATO exhibits significant reward fluctuations initially but rapidly learns offloading strategies through multi-agent collaboration and model aware, eventually converging to approximately $-40$, outperforming all baseline algorithms and demonstrating its efficiency and stability in AIGC task offloading scenarios. In contrast, other algorithms suffer from low decision-making efficiency due to the lack of model aware or multi-agent collaboration. These results demonstrate that MADDPG-MATO excels in model switching and task offloading optimization.

\begin{figure*}[htbp]
    \centering
    \begin{minipage}{0.32\textwidth}
        \centering
        \includegraphics[width=\linewidth]{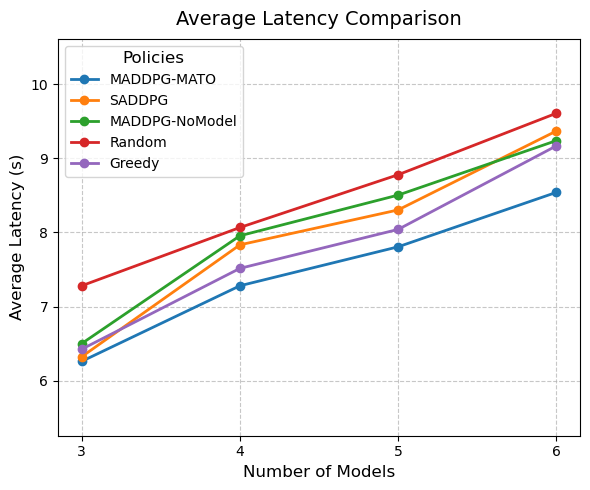}
        \subcaption{ Average Latency VS Models}\label{fig:ed_time}
    \end{minipage}\hspace{0.2em}
    \begin{minipage}{0.32\textwidth}
        \centering
        \includegraphics[width=\linewidth]{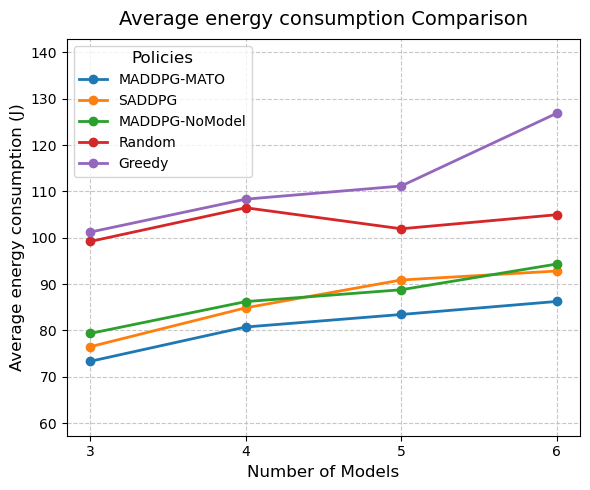}
        \subcaption{Average Energy Consumption VS Models}\label{fig:ed_power}
    \end{minipage}\hspace{0.2em}
    \begin{minipage}{0.32\textwidth}
        \centering
        \includegraphics[width=\linewidth]{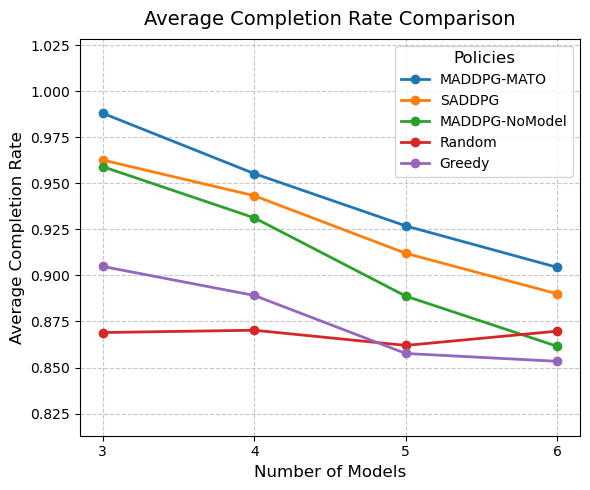}
        \subcaption{Completion Rate VS Models}\label{fig:ed_completion}
    \end{minipage}
    \caption{Performance Comparison Under Different Numbers of Models}
    \label{fig:comparison}
\end{figure*}

\begin{figure*}[htbp]
    \centering
    \begin{minipage}{0.32\textwidth}
        \centering
        \includegraphics[width=\linewidth]{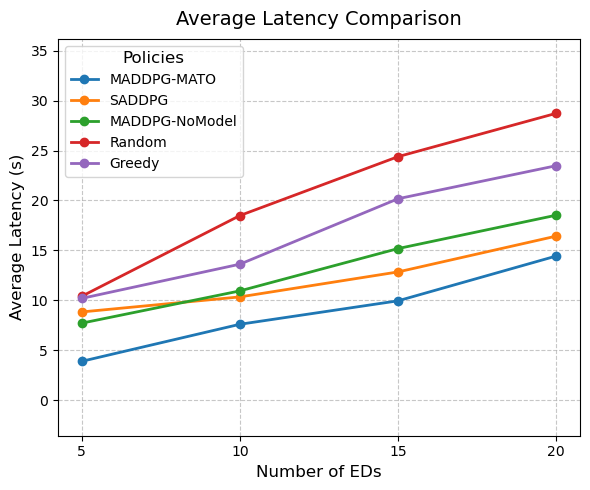}
        \subcaption{ Average Latency VS EDs}\label{fig:ed_time}
    \end{minipage}\hspace{0.2em}
    \begin{minipage}{0.32\textwidth}
        \centering
        \includegraphics[width=\linewidth]{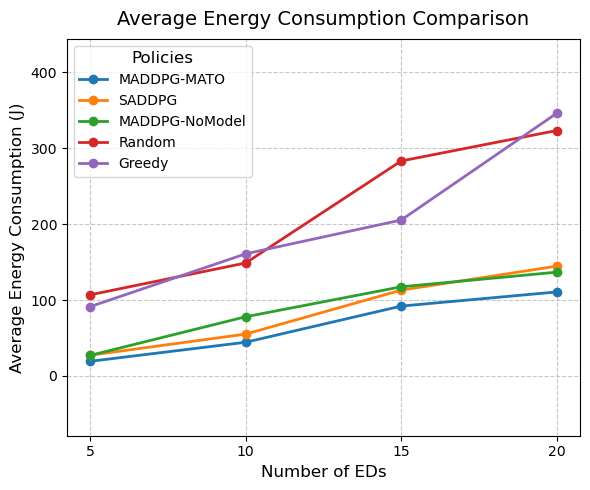}
        \subcaption{Average Energy Consumption VS EDs}\label{fig:ed_power}
    \end{minipage}\hspace{0.2em}
    \begin{minipage}{0.32\textwidth}
        \centering
        \includegraphics[width=\linewidth]{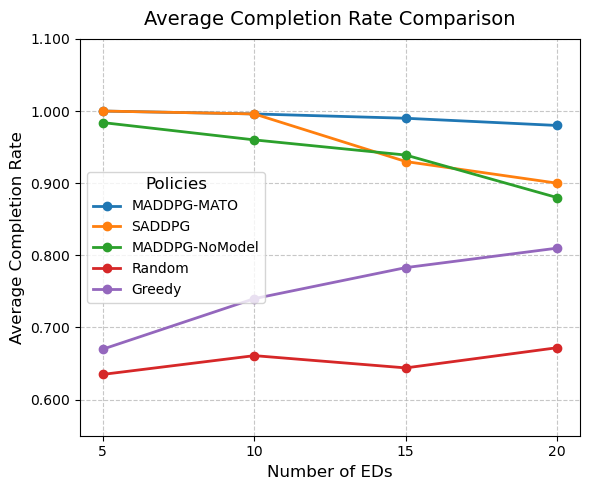}
        \subcaption{Completion Rate VS EDs}\label{fig:ed_completion}
    \end{minipage}
    \caption{Performance Comparison Under Different Numbers of EDs}
    \label{fig:comparison}
\end{figure*}
\subsection{Performance comparison}

To evaluate the effectiveness of MADDPG-MATO optimization in IIoT edge computing, we conducted two sets of experiments to compare the performance of different algorithms in latency, energy consumption, and task completion rate. These experiments respectively investigate the impact of varying the number of AIGC models and the number of EDs on system performance.

\subsubsection{Impact of Different Model Number on System Performance}

AIGC tasks in IIoT are dynamic and diverse, requiring frequent model switching on resource-constrained ES. Higher task diversity increases the frequency of model switching, leading to more frequent dynamic model downloads. Additionally, it is more complicated in dynamic task generation scenario, offloading strategies lacking model aware are easy to offload AIGC tasks to ESs without compatible models and lead to performance decline.

To investigate the impact of task diversity caused by varying model quantities on system performance, we defined a set of AIGC tasks requiring different models. As shown in Fig.3, we increased the number of required AIGC model types from 3 to 6 while keeping other factors constant, evaluating the performance of each algorithm in terms of total latency, energy consumption, and task completion rate.

The results show that with the increase in the number of model types, the frequency of model switching rises, leading to a general increase in latency and energy consumption, while the completion rate generally declining. MADDPG-MATO possesses model aware during offloading decisions, enabling it to make more informed choices in high-diversity scenarios. SADDPG, due to the lack of multi-agent collaboration, experiences a decline in completion rate as the number of models increases,  with higher latency and energy consumption. MADDPG-NoModel ignores model differences, resulting in limited performance, particularly underperforming in high-diversity scenarios. In contrast, MADDPG-MATO significantly enhances performance by optimizing offloading decisions with model awareness. Based on four sets of experimental data with model numbers ranging from 3 to 6, MADDPG-MATO achieves an average reduction of approximately 6.98\% in latency, 7.12\% in energy consumption, and an increase of about 3.72\% in task completion rate. This improvement is particularly notable in high-diversity scenarios, demonstrating the robustness and efficiency of MADDPG-MATO in dynamic task environments.

\subsubsection{Impact of Different Numbers of EDs on System Performance}
An increase in the number of EDs intensifies resource competition and task scheduling complexity, affecting system performance in high-load scenarios. By increasing the number of EDs from 5 to 20, we evaluated the performance of various algorithms in terms of latency, energy consumption, and task completion rate to validate the effectiveness of the proposed algorithm.

As shown in Fig. 4, experimental results demonstrate that MADDPG-MATO employs model awareness to identify compatible ES and leverages multiagent collaboration for efficient task allocation, consistently achieving low latency, reduced energy consumption, and high task completion rates. With 20 EDs, MADDPG-MATO exhibits optimal performance, attaining a task completion rate of 98\%, surpassing other algorithms by at least 11.3\%. In contrast, MADDPG-NoModel, which lacks model awareness and requires frequent model downloads, resulting in higher latency, increased energy consumption, and a task completion rate of approximately 88\%. The Random and Greedy strategies perform the worst, with the highest latency and energy consumption and suboptimal task completion rates.

\section{Conclusion}

This paper introduces a model aware AIGC task offloading algorithm, MADDPG-MATO, tailored for IIoT edge computing to enhance the efficiency of AIGC task offloading. It considers latency and energy consumption induced by model switching into the task offloading strategy. To adapt to the dynamic IIoT environment, EDs are modeled as multiple agents that collaboratively select optimal ES equipped with diverse generative models. Using the MADDPG-MATO algorithm, we derive an optimal model-aware AIGC task offloading decision, significantly enhancing latency, energy efficiency, and task completion rates. Experimental results demonstrate that MADDPG-MATO exhibits superior performance in complex scenarios with increasing model diversity and ED numbers. Through model aware and multi-agent collaboration, it achieves comprehensive optimization of latency, energy consumption, and task completion rate. When model diversity increases, its intelligent switching strategy significantly reduces overhead while maintaining a high completion rate. When the number of EDs increases, MADDPG-MATO effectively optimizes task offloading, maintaining stable performance. These results validate the critical role of model aware and multi-agent collaboration in IIoT edge computing, indicating that MADDPG-MATO is particularly well-suited for AIGC task offloading in dynamic, resource-constrained scenarios.

\bibliographystyle{unsrt}
\bibliography{reference}

\end{document}